\documentclass[12pt]{article}%
\usepackage[T1]{fontenc}
\usepackage[english]{babel}
\usepackage[latin2]{inputenc}
\usepackage{t1enc}
\usepackage{lscape}
\usepackage{amsmath}
\usepackage{amsfonts}
\usepackage{lscape}
\usepackage{indentfirst}
\usepackage{amssymb}
\usepackage{graphicx}%
\usepackage{xcolor}

\begin{document}

\begin{center}
\LARGE \bf {The charm of independent voters}

\large
Ern\H{o} Buz\'as, Attila Szilva \\[6pt]
\end{center}

\vspace{1em}

\begin{abstract}

Independent voters play an increasingly decisive role in contemporary elections, yet their collective behavior remains poorly understood. This paper investigates how a minority of voters with greater flexibility in their political preferences influences opinion formation in polarized electorates. Using a modified Deffuant model, we show that even simple heterogeneity in agents' openness to vote switching can generate rich and unexpected collective outcomes: "open-minded" agents may (i) prevent full convergence into established party blocs, (ii) give rise to transient centrist clusters, or (iii) align with the positions of major parties. These dynamics resemble empirical patterns observed among real-world independent voters. Our results demonstrate that small shifts in openness parameters can substantially reshape the macroscopic structure of political competition, offering a simple explanation for oscillatory electoral outcomes and the emergence of unstable centrist or cross-cutting coalitions.

\end{abstract}

\section{Introduction}

In the past few years a number of Western countries experienced an unusual level of change in political life, which might extend into the next few years. Established parties lost ground and new parties and movements experienced rapid growth. Examples include the rise of Donald Trump and the MAGA movement in the United States, Giorgia Meloni in Italy or the 2024 presidential election shock in Romania \cite{shino2024fall, hartig2025behind, martella2023giorgia, armeanu2025perspectives}. We see events that are hard to interpret for pollsters and political scientists \cite{prosser2018twilight, claassen2025biased, mobasher20252024}. At the same time, it has been found that polls asking respondents about the views of their closest peers can improve prediction accuracy \cite{galesic2018asking}. Moreover, so-called opinion dynamics (OD) models \cite{redner2019reality, sirbu2019algorithmic, shirzadi2025opinion, grabisch2020survey} can accurately predict individual opinions in online populations, as shown in Ref. \cite{vendeville2025voter}. The emergence of these new phenomena may reflect that a growing share of the electorate has become more willing to switch their votes. 

The specific reasons for this willingness can be numerous and quite different, from economic changes in society to demographic, cultural, religious shifts. Corruption scandals and the emergence of new (social) media platforms \cite{PhysRevE.104.044312} can also facilitate the rapid rise of new political actors in the vacuums left by the fall of the previous leaders. Although the overall picture is complex, we focus solely on modeling voters' willingness to change their opinions through interactions with others, which we assume to be a key underlying mechanism \cite{deffuant2000mixing}. While identity-based societal shifts, such as cultural or religious ones are likely to cause systematically biased changes in opinions (such as leaning towards one political side), the behaviour of independent voters whose behavior resembles the Big Five trait of Openness \cite{ramey2021sympathy}, could be really relevant to study the collective effect of "flexible" individuals. Of the Big Five traits, openness and conscientiousness are the most likely to increase the likelihood of vote switching \cite{hamann2025personality}. 

Another relevant category concerns a much larger share of the electorate. These are those who identify as "independents", especially in two-party systems such as the United States of America, where their recently growing share might have become as high as 40-46\%. They do not conform to how partisan voters behave and show significant volatility in their voting preferences over time \cite{reilly2023fluid}. Since a real life independent voter would not stick to a given party or candidate over time, our model captures this behaviour as well. 

While there can be many complex real-life reasons why individual voters depart from established parties and become more susceptible, or more "open-minded", toward extreme or non-mainstream views, we will control the open-mindedness with a single parameter in our computational model, which will be an extension of the Deffuant model \cite{deffuant2000mixing}, a seminal OD model. Note that the adoption of OD as a modeling framework is motivated by empirical findings highlighting the importance of social contagion in political opinions, as reported in Ref. \cite{galesic2018asking}.

The paper is organized as follows. In Section \ref{academic} we present a theoretical overview of OD models, starting  from the classical voter model. In Section \ref{Deffuant} we introduce the basic Deffuant model with random but uniform starting opinion distribution, while in Section \ref{Deffuant-nonuni} the initial distribution of opinion is non-uniform. In Section \ref{variability-open} we add a few agents with higher open-mindedness and study how they change the overall outcome. In Sections \ref{discussion} and \ref{conclusions}, we discuss the meaning of these results and conclude by outlining their possible real-world applications. 

\section{Theoretical background}
\label{academic}

To understand voter behavior, we developed a computational model in the theoretical framework of opinion dynamics. OD is an interdisciplinary field that studies how individuals' opinions evolve through social contagion and other interaction processes, revealing how populations polarize, converge toward consensus, or stabilize in a state of coexistence \cite{redner2019reality, sirbu2019algorithmic, shirzadi2025opinion, grabisch2020survey}. Opinions can be one- or multi-dimensional, binary or continuous, depending on both their object and the way of measurement, and OD deals with their multi-state dynamics on complex networks. The dynamical equations in OD are similar to interaction models known in magnetism \cite{szilva2023quantitative} and statistical physics \cite{castellano2009statistical}.

Historically, the first OD model is the classical voter model (VM) \cite{clifford1973model, holley1975ergodic} in which each individual (voter) can assume one of two discrete states (e.g., yes/no, up/down, Democrat/Republican, $o_{i}=\pm 1$), and a single voter resides at each node of an arbitrary (static) network. A voter at node $i$ is selected at random and copies the state of a randomly chosen neighboring voter ($j$). This dynamics is repeated until consensus is necessarily reached. 

Another seminal early OD model is the DeGroot model in which the opinions of the agents are represented by continuous real numbers and each agent updates their opinion by forming a weighted average of their neighbors' current opinions \cite{degroot1974reaching}. While the classical VM is stochastic and handles the diffusion of categorical labels, the DeGroot model is deterministic and describes the diffusion of continuous signals. Nevertheless, the essence of the models is the same: agents update their states by adopting the states of their neighbors.

Note that the classical VM is one of the few exactly solvable many-body interaction systems when voters are situated on a regular lattice or a homogeneous graph, and it can be shown that the DeGroot model describes the mean-field evolution of the classical VM \cite{grabisch2020survey}. Some models incorporate mobility due to daily commuting, i.e. it is possible to calibrate them to real socio-economic data \cite{fernandez2014voter, michaud2018social}. There are examples when VMs can accurately predict individual opinions in online populations \cite{vendeville2025voter}. A great summary of the binary-state dynamics on complex networks is given by Ref. \cite{gleeson2013binary} and a mini-review about reality inspired VMs has been published by Ref. \cite{redner2019reality}.

While VMs capture social influence, they do not incorporate homophily \cite{thurner2025more}, the well-documented tendency for individuals to connect with others who share similar characteristics or views. One way to take into account homophily is to assume that neighbors on the graph interact at a rate proportional to the number of opinions they have in common. This is the key idea behind the Axelrod model \cite{axelrod1997dissemination} that deals with discrete opinions and multidimensional cultural vector. Another way to consider both homophily and social influence is the Deffuant model \cite{deffuant2000mixing}, a bounded-confidence model, that describes how continuous opinions evolve when individuals interact only with those whose opinions are not too far apart. 

Adding homophily to the models makes them more realistic, however one has to take into account the free will or stubbornness of agents. The Friedkin-Johnsen (FJ) model is a natural extension of the classical DeGroot model in which the agents are trying to remember their original identity. The mathematical formulation of the FJ model is given by Ref. \cite{friedkin1990social}. Our goal is to capture the uniqueness of the voters who are not willing to join the typical large blocks. While the FJ model introduces a parameter to handle the stubbornness the agents, we will modify the classical Deffuant model by introducing inter-individual variability in open-mindedness.

\section{The basic Deffuant opinion dynamics model}
\label{Deffuant}

Consider a population of $n$ agents who belong to an undirected and unweighted social network in which they can be randomly connected. At any given time, an agent $i$ has an opinion $o_{i}$ between $0$ and $1$. Each time step the agents are influenced by a certain set of other agents, and their opinions will be updated due to these interactions. More concretely, agent $i$ will interact with a randomly chosen agent $j$ only if their opinions are close enough to each other. In the model, $i$ and agent $j$ interact with each other only if $|o_{i}-o_{j}| < \epsilon $, where $\epsilon$ is a threshold parameter between $0$ and $1$. Otherwise, they refuse to update their beliefs. The parameter $\epsilon$ measures the open-mindedness of the agents. In addition, we introduce a convergence parameter $\mu$ between $0$ and $0.5$ that measures the strength of the interactions. With these two parameters, the process of opinion updates can be written as follows,
\begin{equation}
o_{i}\left(t+1 \right) = o_{i}\left(t \right) + \mu \left[ o_{j}\left(t\right)-o_{i}\left(t\right)\right] \,,
\label{Deffuant1}
\end{equation}
and
\begin{equation}
o_{j}\left(t+1 \right) = o_{j}\left(t \right) + \mu \left[ o_{i}\left(t\right)-o_{j}\left(t\right)\right] \,,
\label{Deffuant2}
\end{equation}
where $t$ labels the discrete time. 

\begin{figure}[h!]
\centering
\includegraphics[width=1.0\textwidth]{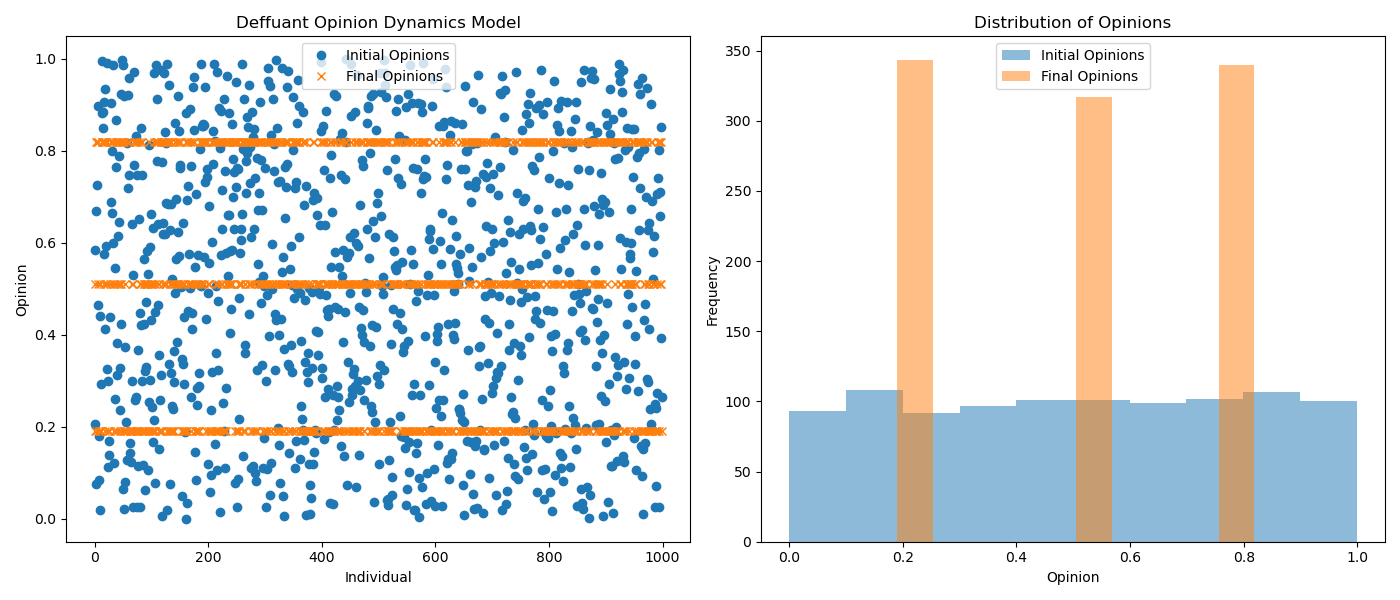}
\par
\vskip -0.5cm  \caption{\footnotesize{Deffuant model with $n=1000$ agents when the parameter $\epsilon$ is $0.15$ and the parameter $\mu$ is $0.2$. On the left panel, the initial opinions of individual agents are represented by blue circles, while the right panel displays their distribution as a histogram with 10 bins. The orange crosses show the final converged state after $t=10^{5}$ iteration steps.}}%
\label{basic_model}%
\end{figure}

Let us take an example in which $n=1000$ agents update their opinions through repeated interactions starting with randomly distributed opinions at $t=0$. Specifically, the initial opinions of individual agents -- represented by blue circles -- are shown in the left panel of Fig. \ref{basic_model}, while the right panel displays their distribution as a histogram with 10 bins. Each blue bin covers around 100 agents since the initial distribution is uniform. Then we set the parameter $\epsilon$ as $0.15$ and the parameter $\mu$ as $0.2$ and run the simulation until $t=100000=10^{5}$. We find that the agents form three clusters as shown by orange in Fig. \ref{basic_model}. This means that three bins cover all agents (right panel) and the agents have the exact same opinions within a cluster (left panel). The qualitative picture will be the same if we repeat the simulation with the same parameter set, only the number of agents in the final bins could sometimes be less uniform. 

\begin{figure}[h!]
\centering
\includegraphics[width=1.0\textwidth]{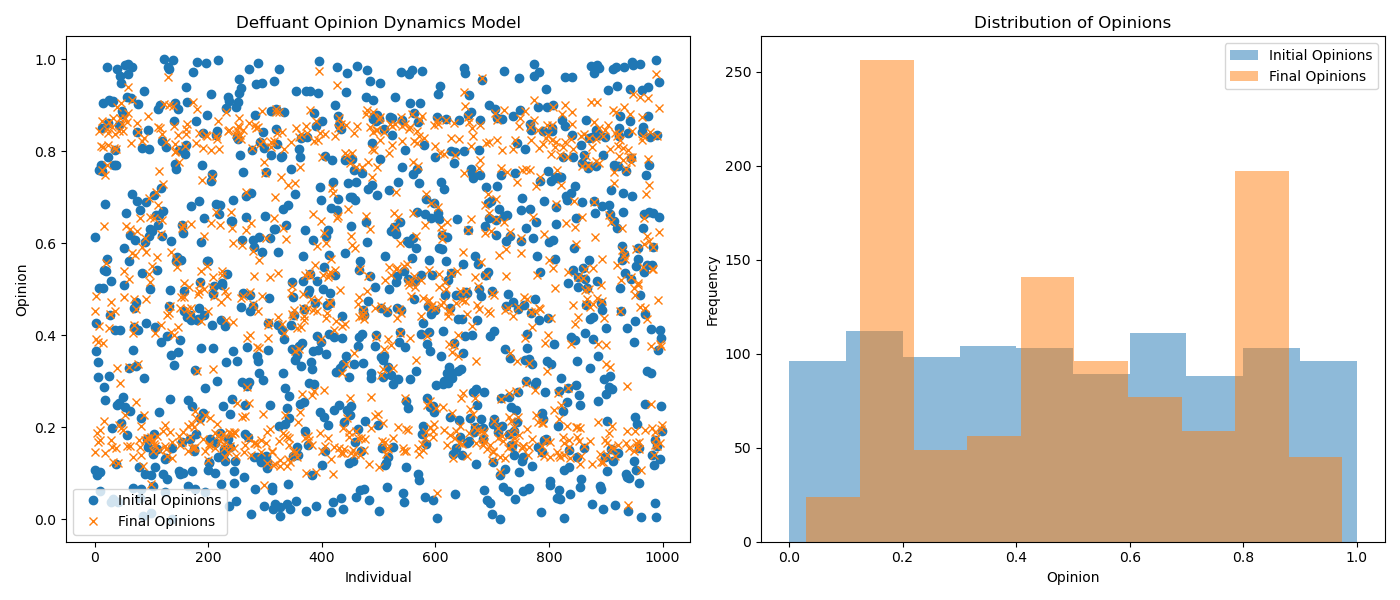}
\par
\vskip -0.5cm  \caption{\footnotesize{A non-converged Deffuant model. All parameter settings are the same as in Fig. \ref{basic_model} expect that the simulation was stopped after $t=20000$ iteration steps.}}%
\label{20k}%
\end{figure}

If we stop the simulation a bit earlier, for example after $t=20000$ iterations, then the final opinions are more scattered as shown in Fig \ref{20k}. However, if we run the simulation ten times longer ($t=10^{6}$) then qualitative picture will not change, i.e., the $t=10^{5}$ case seems a good choice to analyze stable final states. Note that the convergence parameter $\mu$ can be reduced to, say, $0.05$ and then the qualitative picture will be similar to Fig \ref{20k} when we take $10^{5}$ simulation steps. Instead of slowing down the convergence, we will keep $\mu$ at $0.2$ and $n=1000$ (a larger number of agents requires more iterations, too). 

The choice of parameter $\epsilon$ is much more interesting. If $\epsilon > 0.27$ then, in the final state, all opinions will be arranged into a single cluster around $o_{i}=0.5$. If $\epsilon$ is between $0.18$ and $0.27$ then we typically see two opinion clusters in the final stage. The numbers that define these phase boundaries are universal: we get the same boundaries with n=100 or n=10000 agents. Note that decreasing the value of $\epsilon$ leads to the formation of an increasing number of clusters.

\section{Deffuant model with non-uniform starting}
\label{Deffuant-nonuni}

\begin{figure}[h!]
\centering
\includegraphics[width=1.0\textwidth]{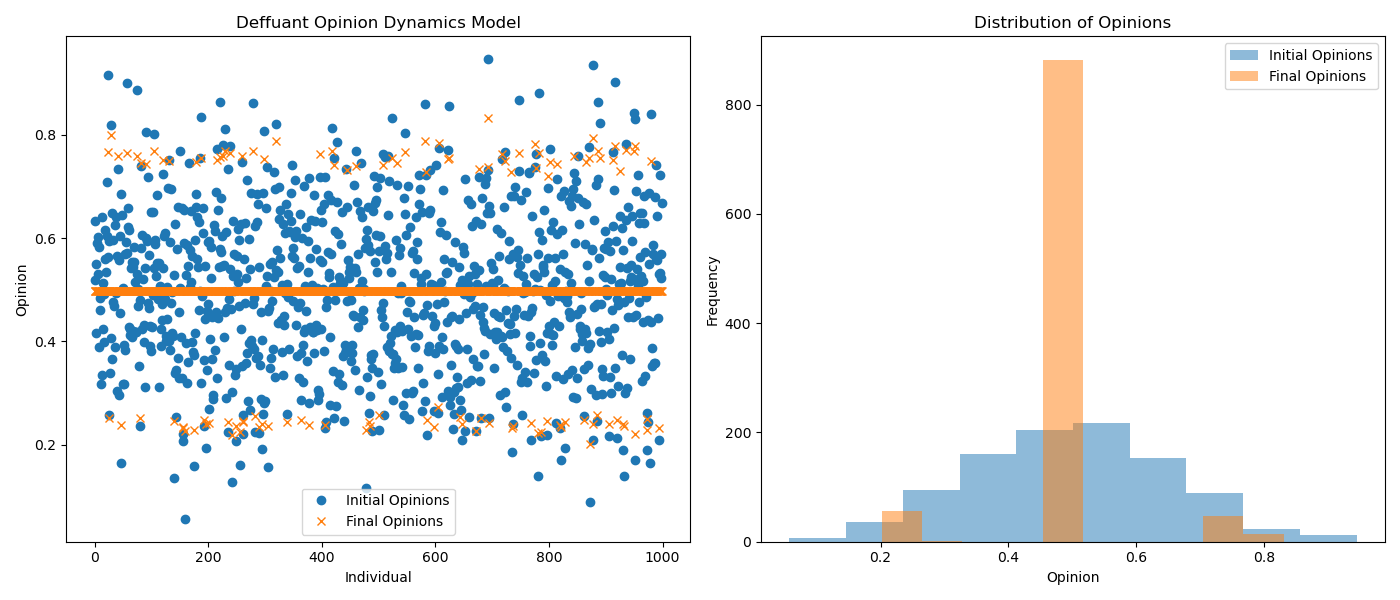}
\par
\vskip -0.5cm  \caption{\footnotesize{Deffuant model with $n=1000$ agents when the parameter $\epsilon$ is $0.15$ and the parameter $\mu$ is $0.2$. However, the initial opinion distribution shown by blue is non-uniform -- instead, it follows a normal distribution with $m=0.5$ mean and $\sigma=0.15$ standard deviation. The final converged states after $t=10^{5}$ iterations are shown by orange.}}%
\label{Gaussian}%
\end{figure}

Let us continue our analysis with the case when the initial opinion distribution is non-uniform. In the first example the initial opinion distribution follows a normal distribution with $m=0.5$ mean and $\sigma=0.15$ standard deviation as shown by blue in Fig. \ref{Gaussian}. Still we can see three clusters in the final stage, however, the vast majority of the agents are located in the middle one. Note that if $\sigma$ is smaller, say, $0.05$ then all agents are arranged in a single cluster around $o_{i}=0.5.$

\begin{figure}[h!]
\centering
\includegraphics[width=1.0\textwidth]{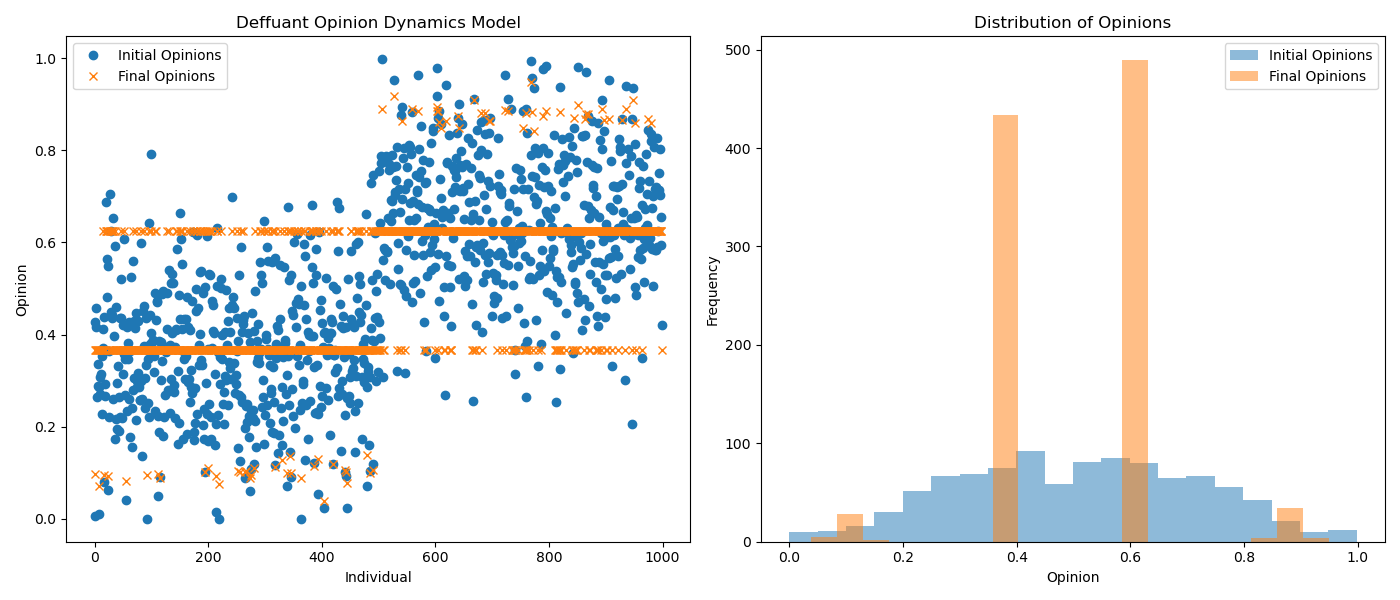}
\par
\vskip -0.5cm  \caption{\footnotesize{Deffuant model with $n=1000$ agents when the parameter $\epsilon$ is $0.15$ and the parameter $\mu$ is $0.2$. However, the initial opinion distribution follows now a bimodal distribution function (a function with two Gaussian peaks) where $m_{1}=0.35$, $m_{2}=0.65$ and $\sigma=0.15$. The final converged states after $t=10^{5}$ iterations are shown by orange.}}%
\label{Bimodal}%
\end{figure}

It can be more realistic if the opinions are polarized around two peaks, hence in the next example we start to run the simulation from a bimodal distribution function (a function with two Gaussian peaks) where $m_{1}=0.35$, $m_{2}=0.65$ and $\sigma=0.15$ as shown by blue in Fig. \ref{Bimodal}. Despite the fact that the parameter $\epsilon$ is still set to $0.15$, we typically get four clusters instead of three. There are two bigger clusters closer to the center (two mainstream parties) and there are two smaller satellite parties on the left and right edges of the spectrum. 

\section{Inter-individual variability in open-mindedness}
\label{variability-open}

As we said, the parameter $\epsilon$ measures the open-mindedness of the agents. As a next step, we introduce heterogeneity in tolerance levels. This can be done in several ways, however, to truly capture the effects of diversity, it seems reasonable to adopt a model that remains simple while incorporating differences in agents' openness to others' opinions.

Hence, we will consider a population of $n = 1000$ agents, where $\nu = 200$ agents have a tolerance level of $2\epsilon$, i.e., twice as large as that of the remaining agents. We refer to these as Open-Minded agents, while the remaining $n - \nu = 800$ agents are referred to as Normal agents. More specifically, when two agents interact, if one of them belongs to the Open-Minded group and their opinion difference is less than $2\epsilon$, the Open-Minded agent will update its opinion according to Eqs. (\ref{Deffuant1})-(\ref{Deffuant2}). Normal agents update their opinions only if the opinion difference is less than $\epsilon$, as in the basic model.

\begin{figure}[h!]
\centering
\includegraphics[width=1.0\textwidth]{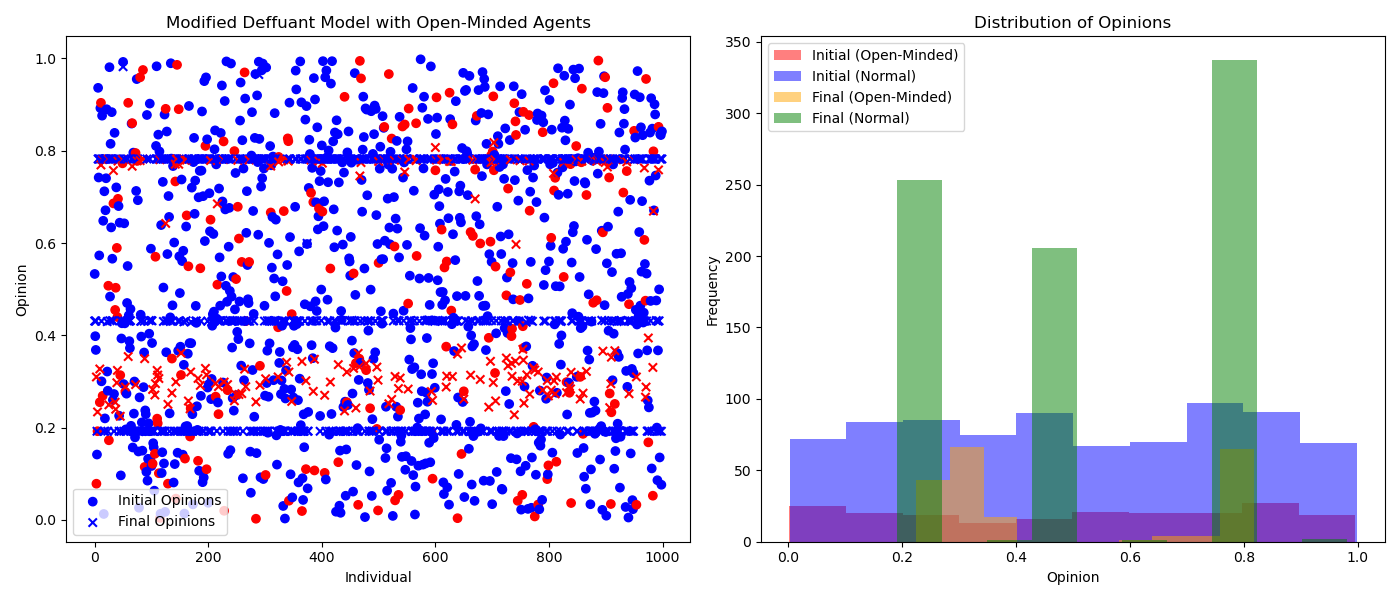}
\par
\vskip -0.5cm  \caption{\footnotesize{Deffuant model with $n=1000$ agents where $\nu = 200$ are Open-Minded when the parameter $\epsilon$ is $0.15$ and the parameter $\mu$ is $0.2$. On the left panel, the initial -- uniform -- opinions of Normal agents are represented by blue circles while the initial -- also uniform -- opinions of Open-Minded agents are represented by red circles. Similarly, the blue and red crosses show the final converged state of the opinions of Normal and Open-Minded agents after $t=10^{5}$ iteration steps, respectively. The right panel displays the distribution of opinions as a histogram (color coding on the top-left corner of the figure).}}%
\label{open_m}%
\end{figure}

The result of the simulation is shown in Fig. \ref{open_m}. On the left panel, we can see that both the Normal and Open-Minded agents' opinions are distributed uniformly shown by blue and red circles, respectively. What is more interesting is that while the Normal agents at final stage (blue crosses) converged to three clusters, the ordering of Open-Minded agents is less pronounced, which can also be seen on the histograms on the right panel of Fig. \ref{open_m}. Note that the qualitative picture will be the same if we run the model for a much longer time -- we have tested it even with $t=10^{8}$ iteration steps. 

\begin{figure}[h!]
\centering
\includegraphics[width=1.0\textwidth]{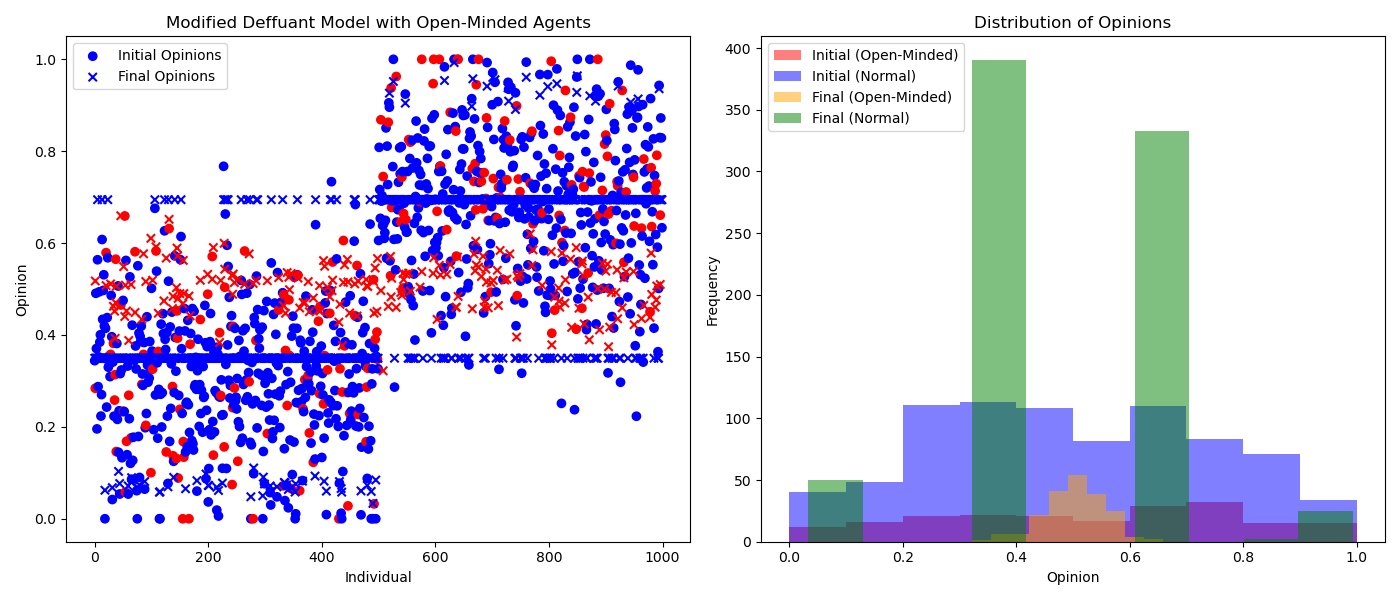}
\par
\vskip -0.5cm  \caption{\footnotesize{Deffuant model with $n=1000$ agents where $\nu = 200$ are Open-Minded when the parameter $\epsilon$ is $0.15$ and the parameter $\mu$ is $0.2$. On the left panel, the initial opinions of Normal agents are represented by blue circles while the initial opinions of Open-Minded agents are represented by red circles. Here, the initial opinion distribution of all agents follows a bimodal distribution function where $m_{1}=0.3$, $m_{2}=0.7$ and $\sigma=0.15$. In this concrete simulation, the Open-Minded voters in the final state are scattered in the center, between the two main parties. The simulation is run for $t = 10^8$, iteration steps.}}
\label{open_bi}%
\end{figure}

\begin{figure}[h!]
\centering
\includegraphics[width=1.0\textwidth]{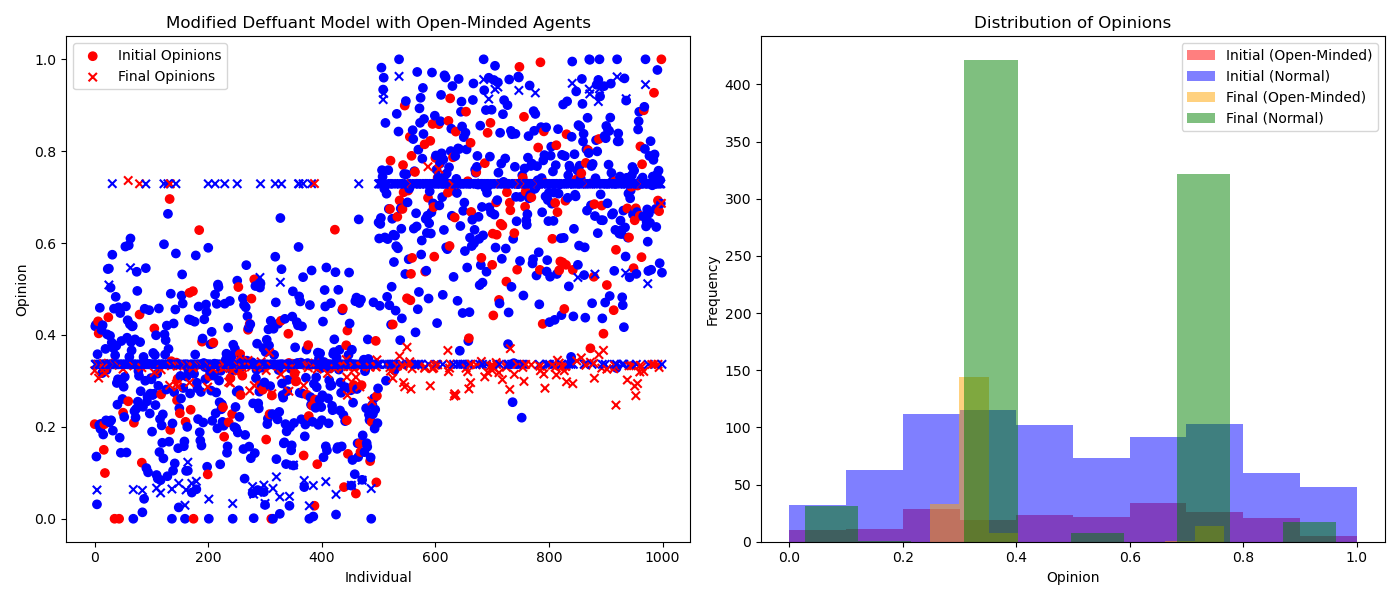}
\par
\vskip -0.5cm  \caption{\footnotesize{Results obtained with the same parameter settings as in Fig. \ref{open_bi}. Open-Minded and Normal agents start from a bimodal initial opinion distribution, and the simulation is run for $t = 10^8$ iterations. In this case, Open-Minded voters in the final state do not converge to the center but instead align with the major parties, or even slightly beyond them toward the extremes of the opinion spectrum.}}%
\label{open_bi2}%
\end{figure}

Fig. \ref{open_bi} and Fig. \ref{open_bi2} show even more interesting cases, where Open-Minded and Normal agents are present in the same way as in Fig. \ref{open_m}, but this time the initial opinion distribution of all agents follows a bimodal distribution, as described in the caption of Fig. \ref{open_bi}. In both figures, we observe the emergence of two main parties -- left and right -- as well as smaller satellite groups at the extremes of the spectrum. However, the Open-Minded voters do not condense into a stable state, even after running the simulation for $t = 10^8$ iteration steps. Instead, they either remain scattered in the center, between the two main parties as in Fig. \ref{open_bi}, or they largely align with the main parties, as seen in Fig. \ref{open_bi2}.

\section{Discussion }
\label{discussion}

We have demonstrated that starting the Deffuant opinion dynamics model simulation with a non-uniform opinion distribution one can get one or two dominant opinion clusters (see Figs. \ref{Gaussian} and \ref{Bimodal}, respectively) in such circumstances when one get three clusters with uniform starting (Fig. \ref{basic_model}). More precisely, in the bimodal initial condition (Fig. \ref{Bimodal}), two major parties emerge in the center-left and center-right regions of the opinion spectrum, while smaller parties form closer to the extremes.

Introducing a 20\% share of "open-minded" voters to a uniform distribution of opinions in the Deffuant opinion dynamics model, we find three competing clusters (Fig. \ref{open_m}) with the previously overbearing center now weaker than its competitors. Under more realistic bimodal initial conditions, these independent voters  either recreate the "centrist" voters, that coalesce in the center, independently of the main parties, as seen in Fig. \ref{open_bi}. Or, alternatively they side with one of the main parties, or even become somewhat more extreme than the main party itself, as it happens in Fig. \ref{open_bi2}.

\section{Conclusions }
\label{conclusions}

Despite relying on simple starting assumptions, our model is sufficiently rich to reproduce the atypical behaviour observed among independent voters. A simple shift in the flexibility of the voters could successfully model the formation of ephemeral centrist blocks in the electorate that are not sticking to their positions as strongly as the rest of the electorate. As shown in the model, targeting them can be enough to sway the outcome without changing anything else in the population. While this is a simple concept in the Deffuant model, the behaviour corresponding to it in real life would appear as a segment of the population getting radicalised in one cycle, then getting either radicalised to the opposite side or getting de-radicalised to a mild standpoint in the centre. This reflects the kind of patterns that have recently proven difficult for pollsters and political scientists to interpret \cite{prosser2018twilight, claassen2025biased, mobasher20252024}. This could also explain why seemingly radical voters of one side would feel compelled by what are seen as radicals on the other side (such as some right-wing MAGA voters siding with left-wing New York mayoral candidate Zohran Mamdani \cite{nbcnews2025underdog}). 

The behavior of independent voters presented in Section \ref{variability-open} captures the characteristic dynamics ("the charm") of open-minded, or independent, voters. In real life setting these states could oscillate, leading to a series of outcomes like the United States of America elections between 2016 and 2025, where important segments of the population felt independent from the main parties and switched their votes, leaning in different directions, weakening traditional voting patterns \cite{shino2024fall, hartig2025behind}. Our model provides a simple, computable explanation for the observed unorthodox behavior and serves as a strong building block for more sophisticated models calibrated with real socio-economic and voting data.

\vspace{2em}

{\large\textbf{Acknowledgement}}

We would like to thank K\'aroly Tak\'acs and Ferenc Sz\H{u}cs for their insightful suggestions.

\bibliographystyle{unsrt}
\bibliography{bibliography.bib}

\end{document}